# Pressure dependent magnetic properties on bulk $CrBr_3$ single crystals

Rubyann Olmos[1], Shamsul Alam[1], Po Hao Chang[1], Kinjal Gandha[2], Ikenna C. Nlebedim[2], Andrew Cole[3], Fazel Tafti[3], Rajendra Zope[1], Srinivasa R. Singamaneni[1*]

[1]Department of Physics, The University of Texas at El Paso, El Paso, TX 79968, USA

[2]Ames Laboratory, Ames, Iowa 50011, USA

[3]Department of Physics, Boston College, Chestnut Hill, MA 02467, USA

* Correspondence: srao@utep.edu

Abstract

The van der Waals class of materials offer an approach to two-dimensional magnetism as their spin fluctuations can be tuned upon exfoliation of layers. Moreover, it has recently been shown that spin-lattice coupling and long-range magnetic ordering can be modified with pressure in van der Waals materials. In this work, the magnetic properties of quasi two-dimensional $CrBr_3$ are reported applying hydrostatic pressure. The application of pressure (0 - 0.8 GPa) shows a 72 % decrease in saturation magnetization with small decrease in the Curie temperature from 33 to 29 K. Density functional theory calculations with pressure up to 1 GPa show a reduction in volume and interplanar distance as pressure increases. To further understand magnetic properties with applied pressure, the magnetocrystalline anisotropy energy (MAE) and exchange coupling parameter (*J*) are calculated. There is minimal decrease in MAE and the first nearest neighbor interaction ($J_1$) ($U$ = 2.7 eV and $J$ = 0.7 eV), shows an increase in $J_1$ with respect to pressure. Overall, $CrBr_3$ displays ferromagnetic interlayer coupling and the calculated exchange coupling and MAE parameters match well with the observations from the experimental work.

1. Introduction

Van der Waals (vdW) materials have proven to be promising candidates for magnetoelectronic, data storage, and memory applications [1-4]. Moreover, bulk magnetoelectrical properties and structural characteristics of the $CrX_3$ (X = Cl, Br, I) and $CrYTe_3$ (Y = Si, Ge) have been amply studied in the literature since the two-dimensional (2D) ferromagnetic (FM) order in $CrI_3$ and $Cr_2Ge_2Te_6$ were first discovered [2,5]. This significant investment placed into the study of vdW materials is by and large due to their ability to retain long-range magnetic ordering down to the low dimensions [1-2, 9-10]. Markedly, the weak interlayer bonding present in vdW crystals have allowed scientists to exfoliate or synthesize these materials to one or few layers despite theoretical limitations regarded in the Mermin-Wagner theorem, forbidding the existence of long-range magnetic ordering in an isotropic Heisenberg system in the two-dimensions [9]. The key component credited with the stabilization of magnetic ordering in vdW magnets is the uniaxial magnetic anisotropy [2,5,10]. The $CrX_3$ family are exfoliable semiconductors with strong magnetic properties stemming from their crystalline structure. In particular, the source of magnetic anisotropy arises from an increase in spin orbit coupling associated with the halogen atom as one moves down the periodic table. Therefore, the magnetocrystalline anisotropy energy (MAE) ascending from the Cr atom and covalently bonded halide atoms counteract thermal agitations allowing magnon excitation gaps to open, thereby, lifting constraints imposed by the Mermin-Wagner theory.

CrBr$_3$ not only exhibits FM ordering in the bulk but as well as in the monolayer limit [5, 11–14]. In the bulk form, CrBr$_3$ exhibits phase transitions from a monoclinic *C2/m* phase at high temperatures to a rhombohedral-trigonal *R3* phase at very low temperatures. The layers in the compound lay along the *ab* plane and stack along the *c* plane where the superexchange mechanism grants strong FM interactions along the easy *c*-axis. In the CrX$_3$ family, magnetic ordering temperatures and Cr-Cr distance increase by the size of halogen atom from Cl to Br to I, which allows for direct exchange to weaken. Furthermore, the Cr-X bond becomes more covalent as electronegativity decreases from Cl to Br to I, leading to a strengthening of the superexchange interactions, in this order. In chromium trihalides, Cr$^{3+}$ ions are arranged in a honeycomb network in edge-sharing octahedral coordination by six X–ions, each bonded to two Cr ions. The resulting slabs of composition CrX$_3$ are stacked with vdW gap separating them.

The application of hydrostatic pressure offers a disorder-less approach in tuning the magnetic and electrical properties of vdW materials. Moreover, when pressure is applied to a vdW system of weakly coupled layers the bond-angle, interlayer coupling, layer separation and stacking order are strongly affected. The Goodenough-Kanamori-Anderson rules state that when the magnetic ion to ligand to magnetic ion angle is $\theta = 90°$, the superexchange interaction is FM. On the other hand, direct exchange is antiferromagnetic (AFM) when $\theta = 180°$ [15, 16]. Manipulation of the superexchange mechanism has been seen in Cr$_2$Ge$_2$Te$_6$ where the Cr-Ge-Cr bond angle diverges from 90 ° when pressure is increased. This tunes the $T_C$ from 66.6 K at 0 GPa to 60.6 K at 1 GPa [17]. However, it is possible for pressure to push the bond angle towards 90° favoring a FM state, a possibility for CrI$_3$ as the Cr-I-Cr bond angle is ~95° [18,19] and similarly for CrBr$_3$ where $\theta = 95.1°$ [13]. Remarkably, CrGeTe$_3$ has recently demonstrated a significant increase in $T_C$ exceeding 250 K for pressures (*P*) greater than 9 GPa, however, this is accompanied by an initial decrease in its $T_C$ in the range $0 < P < \sim 4.5$ GPa [20]. The magnetic phase transition for CrBr$_3$ has been reported to be in the range of 32 to 37 K [12, 21-24]. CrBr$_3$ is relatively stable in air making it easier to work with compared to its other CrX$_3$ family members. Additionally, CrBr$_3$ shows interesting magnetic characteristics such as a temperature dependence of the magnetization and 2D magnetic correlations [25, 26]. Overall, to our knowledge, there are only a couple of works reported on the pressure-induced magnetic properties of the bulk CrBr$_3$ compound [22,23]. While the initial report [22] showed $T_C$ decreases with increase in the pressure, it did not discuss how pressure changes the saturation magnetization. Interestingly, the most recent work [23] reported an increase in the magnetization upon the application of pressure on the same compound. Therefore, to broaden the knowledge and understanding, we employed both experimental and theoretical approaches to study the magnetic properties of CrBr$_3$ with pressures up to 1 GPa.

2. Methodology

    2.1 Experimental

Single crystal samples of CrBr$_3$ were prepared using chemical vapor transport by placing powders of Cr metal and TeBr$_4$ inside a quartz tube and maintaining the hot and cold zones at 700 and 600 °C for 5 days. The green plate-like crystals are a few millimeters across and 50 microns thick [27,28]. Magnetic measurements were performed using a Quantum Design MPMS 3 SQUID magnetometer. Isothermal magnetization measurements were taken at 10 K with a ± 3 T magnetic field. Zero-field cool (ZFC) temperature dependent magnetization measurements were performed from 2 - 400 K. Hydrostatic pressure was applied using a Quantum Design piston cell. The pressure transmitting medium was Daphne oil and a Pb manometer was used to monitor the pressure in the cell. Compression on the cell was increasingly applied and only depressurized after the data collection was completed.

2.2 Computational Details

The Density functional theory (DFT) calculations of vdW layered $CrBr_3$ were carried out with Vienna ab initio (VASP) code [29,30] within projector augmented-wave (PAW) method [31,32]. General gradient approximation (GGA) in the Perdew-Burke-Enzerhof (PBE) [33] were used for the exchange correlation functional. The plane-wave cut-off energy is 500 eV and an 8×8×3 $k$-point is used to sample the Brillouin zone [34]. For the relaxation of the crystal structure, the non-local vdW functional in form of optB88-vdW [35,36] is incorporated to account for the interlayer vdW force, and all the lattice constants and ionic coordinates were relaxed until the maximum force on all ions is less than $5 \times 10^{-3}$ meV/Å. For the magnetic property calculations, the on-site Coulomb interactions are taken into account using LDA+$U$ [37] to improve the description of the interactions between localized $d$ electrons of transition atoms. The hydrostatic pressure effect calculation was done by adding the PSTRESS [38] tag, which adds the stress to the stress tensor and an energy generated from the external pressure.

3. Results and Discussion

In Fig. 1(a), we present the isothermal magnetization at 10 K collected at various pressures up to 0.8 GPa. A dramatic decrease in magnetization is seen immediately upon the smallest application of pressure (0.2 GPa). The saturation magnetization ($M_s$) at 0 GPa transitions from 21.09 to 5.86 emu/g at 0.8 GPa, over a 72.20 % decrease. Fig. 1(b) displays the temperature dependence of magnetization collected at various pressures. The pressure-dependent magnetization extracted from Fig. 1(b) follows the same trend as obtained from the isothermal magnetization, Fig. 1(a), where minor compression of the cell results in a decrease in magnetization. It should be noted that similar values of $M_s$ are seen for pressures in the range 0.2 - 0.6 GPa as shown in Fig. 1(c) where the initial drop in magnetization is seen leveling off at about 6.0 emu/g. The Curie temperature ($T_C$) is extracted from the derivative (d$M$/d$T$) of the temperature dependent magnetization data. The variation of $T_C$ as a function of pressure is plotted in Fig. 1(c), which is monotonically decreasing although by only ~ 4 K at the largest pressure. The decrease in $T_C$ and in magnetization is presumably due to pressure affecting the overall FM interactions in the system. Moreover, if pressure causes the bond angle to deviate away from a 90° for the Cr-X-Cr bond angle, $T_C$ is expected to decrease—weakening the (FM) superexchange interaction [15,16].

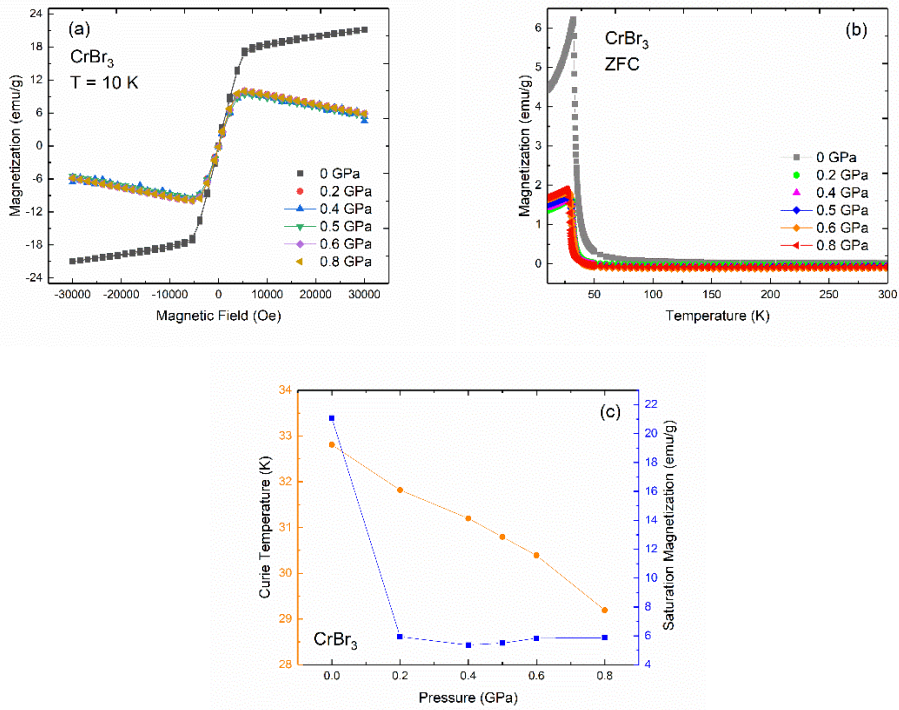

Figure 1 (a) Isothermal magnetization for pressure up to 1 GPa at 10 K in the ferromagnetic region with ± 3 T magnetic field. (b) Zero field cool temperature dependent magnetization measurements from 2 - 400 K (showing up to 300 K) for 0 - 0.8 GPa. (c) Curie temperature, $T_C$, (left, orange) and saturation magnetization, $M_s$, (right, blue) plotted as a function of pressure.

To gain further insights on the pressure effect within the 1 GPa range, we performed first-principles calculations examining the structure of CrBr$_3$ and its magnetic properties. On-site Coulomb interactions are particularly strong for the localized *d* electrons in the CrBr$_3$ system. To remedy this shortcoming in this correlated system the Hubbard-*U* method pioneered by Anisimov *et al*. is applied [37]. In treating the localized *d*-electron states from Cr, we use parameters $U = 2.7$ eV and $J = 0.7$ eV. Fig. 2(a) shows the pressure dependence of the lattice ratio *c*/*a* as a function of pressure showing a monotonic decrease up to 1 GPa. Similarly, the volume of the crystal is decreasing with increasing pressure, as seen in Fig. 5(b). This suggests that the pressure is compressing the interatomic layer separation of the lattice. It is not surprising that the reduction of *c* is more significant due to weak interlayer coupling. A similar trend is also observed in several other vdW layered systems such as Cr$_2$Si$_2$Te$_6$ [17].

To understand how the magnetic properties vary with pressure the exchange coupling *J* and MAE are considered. As discussed in the previous section, the external pressure changes the lattice constants which influence the inter-site electron hopping process. Two of the main mechanisms that contribute to exchange coupling between the localized moments are direct and super-exchange interactions. It is often the result of the competition between the two that dictates the response of the applied pressure. In our model, we consider only the (intra-plane) nearest neighbor exchange interaction $J_1$. The super-exchange interaction, due to the presence of the non-magnetic Br atom in between the Cr ions, increases with applied pressure and subsequently, $J_1$ increases with the isothermal compressibility, see Fig. 2(c). This interaction originates due to the virtual hopping of electrons between the two nearest neighbor Cr-ions via the Br ion. In Fig. 2(c), a small but noticeable change due to the applied pressure is seen where $J_1$ increases almost linearly for $P <$ 0.4 GPa while increments are very minimal for $0.5 < P < 1.0$ GPa.

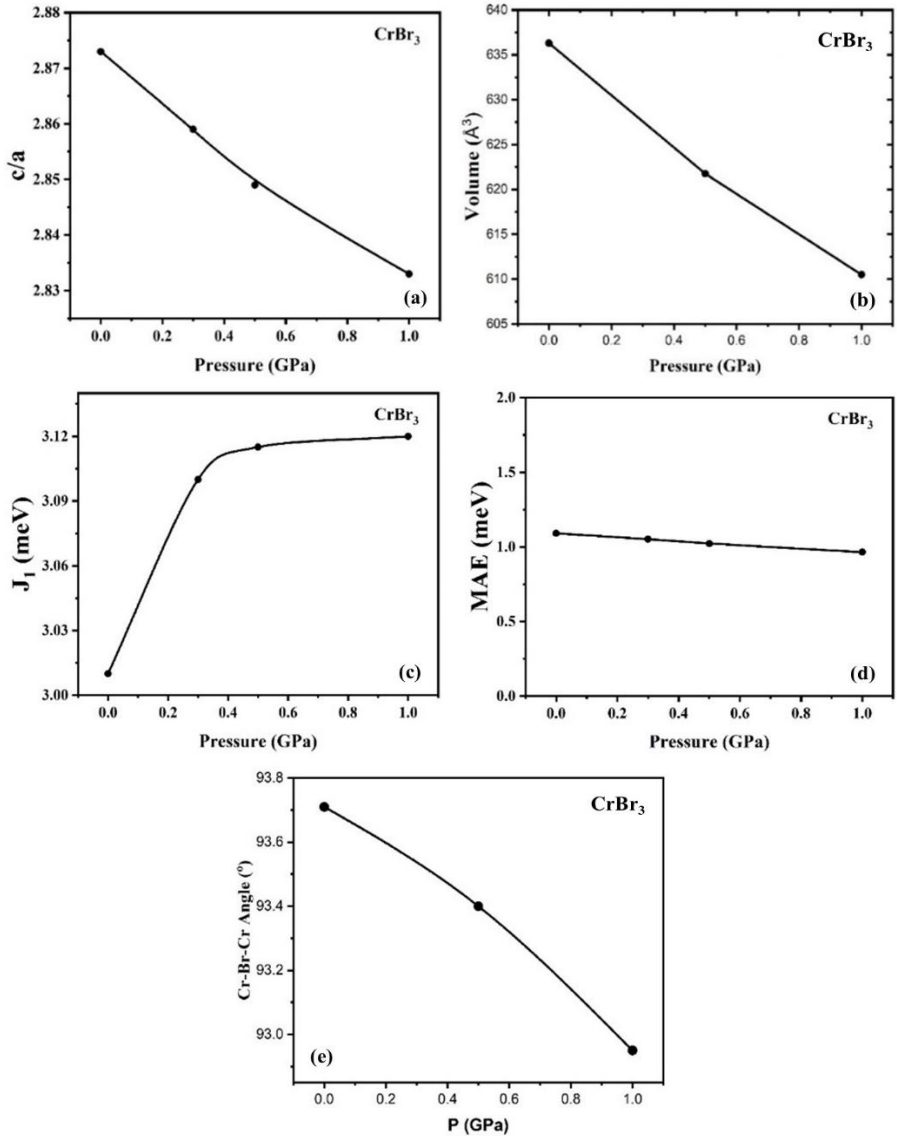

Figure 2(a) Lattice parameter ratio $c/a$ and the (b) volume change plotted versus pressure. (c) First nearest neighbor exchange coupling up to 1 GPa. (d) Magnetocrystalline anisotropy energy as a function of hydrostatic pressure. (e) Reduction of the Cr-Br-Cr bond angle as a function of pressure.

Experimentally [22], a previous report on $CrBr_3$ showed that the $T_C$ decreases from 35.2 K (0 GPa) to about 33 K (1.2 GPa) as represented in its negative pressure coefficient $(dT_C/dP) = -0.2$ K/kbar, implying a negative dependence of $J_1$ with pressure. From the theoretical perspective, the FM in-plane super-exchange interaction between the $Cr^{3+}$ in bulk $CrBr_3$ appears to be more dominant than the direct exchange (AFM) interaction which leads to the increase of $J_1$ with pressure in our calculation. In the experiment, H. Yoshida claims that the exchange interaction becomes stronger with decreasing atomic distance as there are stronger orbital overlaps [22]. For example, Br has a smaller atomic radius (114 pm) than that of its neighbor I (133 pm), which can contribute to the dominant Cr-Cr direct exchange interaction. Combining all these effects, the competition between the direct exchange and the indirect super-exchange interaction determines the nature of the $J_1$ dependence with pressure for $CrBr_3$. The competition between AFM direct exchange through Cr–Cr bonding, and FM super-exchange through Cr–Br–Cr bonding, are what determine the nature of the $J_1$ dependence with pressure for $CrBr_3$.

The pressure dependence of MAE with the inclusion of spin orbit coupling is shown in Fig. 2(d) where no significant change in MAE with respect to pressure is observed. When comparing $CrBr_3$ to $CrI_3$, for example, the Cr-Cr separation increases with expanding halogen size from Br to I, as a result the direct exchange (overlap between neighboring Cr orbitals) weakens which successively, enhances the covalent nature of Cr-X bond [39]. Though decrease in MAE contributes to the decrease in $T_C$, it is important to note that the change in bond angle, Fig. 2(e), and the increase in interlayer coupling, Fig. 2(c), could give rise to larger $T_C$ if the system goes to higher pressures. Such an interesting case has recently been seen for $CrGeTe_3$ where $T_C$ initially decreases with increasing pressure, then significantly increases above 250 K at pressures above 9.1 GPa [20]. Therefore, the relationship of MAE and $T_C$ might not be consistent with the general trend observed in $CrBr_3$ and can warrant further investigations into the magnetic behavior at pressures greater than 1 GPa.

4. Conclusion

In this work, we explored the magnetic characteristics of the less studied chromium halide, $CrBr_3$, with pressure as the tuning parameter. We reveal through experiment that $M_s$ and $T_C$ both decrease upon increasing the pressure up to 0.8 GPa. Moreover, through computations we realize that the role of pressure on $J$ values is very complicated and gives rise to further questioning, especially considering that there are a very few pressure-induced calculations and experiments performed on the bulk $CrBr_3$ system. Additionally, the behavior of MAE as function of increasing pressure directly relates to the decrease in $T_C$ our work. Although, $T_C$ was not drastically altered for the bulk $CrBr_3$ case, recent works have shown the importance of carrying out pressure dependent studies at significantly high pressures, thus, it is important to fully characterize the pressure dependent magnetic properties of $CrBr_3$ at pressures above 1 GPa in the future.

CRediT authorship contribution statement

Rubyann Olmos: Conceptualization, Data curation, Investigation, Writing – original draft, Writing – review & editing. Shamsul Alam: Software, Investigation, Data curation. Po Hao Chang: Software, Data curation, Writing – review & editing. Kinjal Gandha: Data curation, Resources. Ikenna C. Nlebedim: Data curation, Resources. Andrew Cole: Resources. Fazel Tafti: Resources. Rajendra Zope: Software, Writing – review & editing. Srinivasa R. Singamaneni: Supervision, Writing – review & editing.

Declaration of Competing Interest

The authors declare that they have no known competing financial interests or personal relationships that could have appeared to influence the work reported in this paper.

Acknowledgments

This material is based upon work supported by the National Science Foundation Graduate Research Fellowship Program under Grant No. 1848741. Any opinions, findings, and conclusions or recommendations expressed in this material are those of the author(s) and do not necessarily reflect the views of the National Science Foundation. S.R.S. and R.O. acknowledge support from the NSF-DMR (Award No. 2105109). SRS acknowledges support from NSF-MRI (Award No. 2018067). PHC and RZ are supported by the US Department of Energy, Office of Science, Office of Basic Energy Sciences, as part of the Computational Chemical Sciences Program under Award No. DE-SC0018331. Support for computational time at the Texas Advanced Computing Center through NSF Grant No. TG-DMR090071.

Conflict of Interest Statement

The authors declare no conflict of interest.